\documentclass[aps,prl,reprint,superscriptaddress,nofootinbib,floatfix]{revtex4-2}

\usepackage{amsmath,amssymb,bm}
\usepackage{booktabs}
\usepackage{hyperref}
\usepackage{tikz}
\usetikzlibrary{arrows.meta,positioning}

\hypersetup{colorlinks=true,linkcolor=blue,citecolor=blue,urlcolor=blue}

\newcommand{\ket}[1]{|#1\rangle}
\newcommand{\Enc}{\mathrm{Enc}}
\newcommand{\Good}{\mathcal G}
\newcommand{\Pairs}{\mathcal D}

\begin{document}

\title{Structured Variational Key Concentration for a Reduced AES Cipher}

\author{Xi Li}
\affiliation{Henan University, Kaifeng, China}

\date{\today}

\begin{abstract}
Grover search gives the generic quantum baseline for symmetric-key recovery by amplifying a reversible public-verification oracle. We investigate whether public round structure can instead shape a variational key register before exact verification. For simplified AES (S-AES), we construct phase separators from S-box differential-distribution constraints, candidate-key-dependent inverse diagnostics, and a global residual Hamming-shell term; each has a reversible compute--phase--uncompute realization. We prove a one-sided completeness statement for these diagnostics and derive a conditional lower-tail bound for false keys under a weakly correlated diagnostic model. On ten random three-pair 16-bit S-AES instances, an 18-layer schedule optimized with Adam attains a mean public-consistent-key probability of $46.78\%$, ranks the correct key first in every run, and limits the largest observed false-key probability to $1.11\%$. The 16-bit study evaluates the induced key-register dynamics without shot sampling, while a separate gate-level S-DES experiment validates the reversible oracle pattern without a key-space phase table. The results establish a controlled reduced-cipher benchmark; they do not claim an attack on full AES or an asymptotic advantage over Grover. Within these limits, the construction provides a useful framework for analyzing how AES structure can be incorporated into quantum algorithms.
\end{abstract}

\maketitle

\textit{Problem setting and scope.---}
Quantum cryptanalysis of a block cipher has a well-defined generic construction: implement the encryption map reversibly, mark keys satisfying $\Enc_k(P_j)=C_j$, and apply amplitude amplification~\cite{Grover1996,GrasslAES,BonnetainAES}. Variational algorithms such as QAOA~\cite{FarhiQAOA,HadfieldAOA,CerezoVQA} permit a different design choice. Public round structure can be used to define intermediate phase separators, so that the key register is shaped by cipher-derived information before exact public verification. The question addressed here is whether this construction produces measurable key concentration in a reduced AES model.

This distinction matters for the success metric. In several AES-like VQAA studies, including studies of S-DES, the known ciphertext is encoded as the target of a direct Hamiltonian~\cite{WangVQAA,AizpuruaVQAA,AizpuruaTN}. Such an objective is well defined, but overlap with a ciphertext state is not, by itself, a unique-key recovery probability: a reduced cipher and a single public pair may admit several consistent keys. We therefore measure probability on the key register and define success with respect to the full public-consistent key set.

We construct the key-register ansatz from cipher-native diagnostics rather than from an arbitrary ciphertext target. The phase family combines S-box differential likelihoods, inverse-round residuals, residual Hamming shells, and one-pair number-partitioning-inspired (NPP) anchors~\cite{MertensNPP}. The latter label describes the geometry of the auxiliary anchor, not a formal reduction from number partitioning. The number of public pairs is a control parameter: one pair does not provide a pairwise DDT constraint, two pairs provide one such constraint, and three pairs provide the three pairwise differences used in the principal benchmark. This permits the effect of additional public information to be studied without changing the underlying circuit construction.

The evidence is separated by role. Gate-level circuits establish the reversible form of the diagnostic oracles; the 16-bit S-AES study quantifies key-register concentration in a fully auditable reduced model; and the resource analysis identifies the width and gate costs that govern a larger implementation. A faithful state-vector simulation must include the cipher-state, key-schedule, diagnostic, and counter workspaces in addition to the key register; these work registers are returned to zero by uncomputation.

\textit{Reduced cipher and reversible modules.---}
The cipher studied below is the standard 16-bit simplified AES used as a reduced AES testbed~\cite{NISTAES,DaemenRijmen,SchaeferSDES,Stallings}. A 16-bit state is four nibbles arranged as a $2\times2$ state; the public 4-bit S-box is
\[
    S=(9,4,A,B,D,1,8,5,6,2,0,3,C,E,F,7).
\]
The 16-bit master key expands to three round keys $K_0,K_1,K_2$. Encryption is
\begin{align}
    S_0 &= P\oplus K_0,\\
    S_1 &= {\rm MC}({\rm SR}({\rm SN}(S_0)))\oplus K_1,\\
    C &= {\rm SR}({\rm SN}(S_1))\oplus K_2,
    \label{eq:saes-main}
\end{align}
where SN applies the S-box to all four nibbles, SR swaps the second-row nibbles, and MC applies the usual S-AES MixColumns over $\mathrm{GF}(2^4)$. The corresponding quantum circuit modules are reversible permutations: AddRoundKey is XOR by a round-key register, SubNibbles is four 4-qubit S-box permutation gates, ShiftRows is swaps, MixColumns is two reversible 8-qubit linear maps, and key expansion is a reversible byte update. These modules are exactly what the attack uses: DDT phases use the final-round S-box and key-xor cancellation, while residual phases use inverse-round versions of the same modules. A larger module-to-Hamiltonian diagram is given in the Supplement.

\textit{Public-consistent key probability.---}
For public data
\begin{equation}
    \Pairs=\{(P_j,C_j)\}_{j=1}^{m},
\end{equation}
the relevant key-recovery set is
\begin{equation}
    \Good=\{k:\Enc_k(P_j)=C_j,\ j=1,\ldots,m\}.
\end{equation}
The metric reported here is the direct key-register probability
\begin{equation}
    p_{\Good}(\Theta)=\sum_{k\in\Good}|\langle k|\psi(\Theta)\rangle|^2.
    \label{eq:pgood}
\end{equation}
Equation~(\ref{eq:pgood}) is an evaluation metric, not an input, oracle, or training label. The set $\Good$ is defined by the public plaintext--ciphertext equations alone. In the reduced 16-bit benchmark, the sum is evaluated exactly after state preparation to remove shot noise; a physical implementation would instead sample the key register and verify only the observed candidates using the same public equations. When $\Good$ is a singleton, this is the true-key probability. This differs from a ciphertext-register ground-state probability unless the public data uniquely identify the key and the state is read out in the key register.

\textit{Key-register construction.---}
The input to the recovery procedure is the public cipher specification and the set of plaintext--ciphertext pairs $\Pairs$. The key register is initialized in $\ket{+}^{\otimes n}$. Each problem layer computes a public diagnostic from a candidate key, applies a phase determined by the diagnostic, uncomputes the associated work registers, and then applies a mixer to the key register. Component diagnostics within a layer are composed serially. After pullback to the key register, their phase operators are diagonal and therefore compose as
\[
    \prod_{r\in\mathcal R_\ell} e^{-i\gamma_{\ell r} H_{\ell,r}(k)}
    = e^{-i\sum_{r\in\mathcal R_\ell}\gamma_{\ell r} H_{\ell,r}(k)}.
\]
Here $\mathcal R_\ell$ is the set of diagnostic components used in layer $\ell$, and the component angles $\gamma_{\ell r}$ are trainable. The phases are therefore generated by unitary composition, not by direct modification of amplitudes. After angle optimization, the key register is sampled and the measured candidates are checked with the public predicate in Eq.~(\ref{eq:pgood}). At the circuit level, the construction does not enumerate the key space or load a length-$2^n$ phase table; each phase is implemented by a reversible compute--phase--uncompute sandwich. The compressed key-register representation used in the 16-bit study evaluates the induced action of these sandwiches exactly for the reduced benchmark.

For the 16-bit benchmark, the pulled-back phases are stored as exact key-register arrays and the small key space is summed exactly when evaluating the objective. This removes sampling noise from the reduced-model comparison; it is not a phase-loading prescription for the scalable circuit. In a gate-level implementation, objective estimates would be obtained from repeated circuit executions, and only measured key candidates would be checked against the public ciphertexts. The DDT, residual, and GHD Hamiltonians are functions of diagnostics computed from the candidate key in superposition. The secret key is used only to generate synthetic benchmark ciphertexts and is not supplied to the training objective.

\textit{Problem Hamiltonians.---}
The first Hamiltonian is a DDT preparation term. For two public pairs $a,b$, the last S-AES key addition cancels in the ciphertext difference, giving a public final-S-box output difference
\begin{equation}
    \Delta y_{ab}=\mathrm{SR}^{-1}(C_a\oplus C_b).
\end{equation}
For candidate key $k$, the circuit computes the corresponding final-S-box input difference $\Delta x_{ab}(k)$. The DDT phase assigns each nibble an energy
\begin{equation}
    h_{\rm DDT}(u,v)=
    \begin{cases}
    -\log[(N_{uv}+\epsilon)/16],&N_{uv}>0,\\
    \zeta,&N_{uv}=0,
\end{cases}
\label{eq:ddt}
\end{equation}
where $N_{uv}=\mathrm{DDT}(u,v)$. The reported runs use $\epsilon=0$ and $\zeta=6$; a positive $\epsilon$ may be used as a numerical regularizer without changing the diagnostic construction. Low energy means that the candidate key explains the observed public difference through likely S-box transitions. This phase is useful for concentration, but it is not a unique-key selector; differential-equivalent false keys can receive the same score.

The second Hamiltonian is a smooth residual term. For public pair $j$ and checkpoint $t$, define the synchronized residual
\begin{equation}
    \delta_{j,t}(k)
    =
    R_{t,k}(\Enc_k(P_j))\oplus R_{t,k}(C_j),
    \label{eq:residual}
\end{equation}
where $R_{t,k}$ is a candidate-key-dependent reversible diagnostic map compiled from the public cipher and key schedule. The same map is applied to the candidate encryption branch and to the public ciphertext branch. A public-consistent key gives $\delta_{j,t}(k)=0$ for every selected diagnostic. The smooth residual phase uses nibblewise Hamming weights of $\delta_{j,t}(k)$, so its physical meaning is local inverse-diagnostic agreement rather than final ciphertext overlap.

The third Hamiltonian is the global residual Hamming-distance (GHD) shell anchor. Let $b$ be the block size ($b=16$ for S-AES) and define
\begin{align}
    d_{j,t}(k)&=\min\{|\delta_{j,t}(k)|_1,\ b-|\delta_{j,t}(k)|_1\},\\
    d(k)&=\sum_{j,t}d_{j,t}(k).
\end{align}
The complement branch preserves the standard planted-NPP complement symmetry used in the anchor construction. Writing $d=d(k)$, the global shell applies a single component
\begin{equation}
H_{\rm GHD}(k)=
\begin{cases}
-\alpha_0,& d=0,\\
    -\alpha_1(r_{\rm sh}+1-d)/(r_{\rm sh}+1),&0<d\le r_{\rm sh},\\
    \eta d,&d>r_{\rm sh}.
\end{cases}
\label{eq:ghd}
\end{equation}
The exact public-consistent branch has $d=0$. Because the complement branch is retained, complement-equivalent residual signatures can also lie at $d=0$; final public verification removes this degeneracy. Near signatures receive a softer attractive phase, and distant signatures are penalized. Here $r_{\rm sh}$ is the integer shell radius used by the implementation. ``Global'' means that all public pairs and all selected checkpoints are aggregated into one residual-distance component; it does not mean all-to-all interactions between arbitrary qubits.

The staged QAOA state is
\begin{equation}
    \begin{split}
    \ket{\psi(\Theta)}
    ={}&
    \prod_{\ell=1}^{p}
    e^{-i\beta_\ell(\sum_i X_i+b_\ell\sum_i n_i)}
    \\[-0.15em]
    &{}\times
    \exp\!\left[
        -i\sum_{r\in\mathcal R_\ell}
        \gamma_{\ell r}H_{\ell,r}
    \right]
    \ket{+}^{\otimes n},
    \end{split}
    \label{eq:qaoa}
\end{equation}
where $\mathcal R_\ell$ specifies the DDT, residual, or global Hamming-shell components used in layer $\ell$. The scalar $b_\ell$ is a trainable layer-wise mixer bias. Setting $b_\ell=0$ recovers the standard transverse-field mixer.

\textit{Diagnostic-separation model.---}
The phase construction admits a simple statistical interpretation. Let
\begin{equation}
    E(k)=\sum_{r=1}^{L} e_r(k)
\end{equation}
be the shifted energy obtained from the selected public diagnostics, with the public-consistent reference subtracted. In the weakly correlated diagnostic model detailed in the Supplement, false-key component gaps have positive mean and controlled lower tails. Then for any low-energy threshold $\tau$ below the mean gap,
\begin{equation}
    \mathbb E\,N_{\rm false}(\tau)
    \le
    (2^n-|\Good|)\exp[-L I(\tau)],
    \label{eq:false-count}
\end{equation}
for a positive rate function $I(\tau)$. Thus, under the stated model, the expected number of false keys in a low-energy band decreases exponentially with the number of informative diagnostic components. The dependence on public data follows directly: all pairwise DDT differences give $4\binom{m}{2}$ components in 16-bit S-AES, while residual and GHD diagnostics scale as $m|\mathcal T|$.

The residual/GHD part gives a concrete instance of this model. If the raw residual bits of a false key are approximately unbiased, then for $B$ aggregated residual bits,
\begin{equation}
    \Pr[w(k)\le \rho B]
    \le
    2^{-B[1-H_2(\rho)]},
    \label{eq:hamming-tail}
\end{equation}
where $w(k)=|\delta(k)|_1$ is the raw Hamming weight and $H_2$ is the binary entropy. A public-consistent key has $w=0$ and hence also $d=0$ after complement minimization, whereas false keys enter a small residual Hamming shell with exponentially small probability under the stated model. The Supplement gives the full proof, the DDT likelihood version, and the circuit-cost statement for realizing $E(k)$ without loading a key table.

The connection from landscape separation to a trained state can also be stated without assuming that finite-depth QAOA always finds the optimum. Let $\ket{\phi_\eta}$ denote the ideal energy filter defined in the Supplement and let $\Pi_{\Good}$ be the projector onto the public-consistent key space. If the trained state obeys $\|\ket{\psi(\Theta)}-\ket{\phi_\eta}\|_2\le\varepsilon$, then
\begin{equation}
    p_{\Good}(\Theta)
    \ge
    \left[\max\left\{0,\sqrt{p_{\Good}(\phi_\eta)}-\varepsilon\right\}\right]^2.
    \label{eq:qaoa-transfer}
\end{equation}
The diagnostic-separation bound controls $p_{\Good}(\phi_\eta)$ through the energy landscape, while Eq.~(\ref{eq:qaoa-transfer}) isolates the finite-depth optimization error. That error is an empirical quantity for the cipher-derived Hamiltonian and is therefore reported through repeated circuit evaluations.

Under the weak-correlation model, the proposition establishes separation of the public-consistent band from the false-key lower tail. It does not provide a finite-depth QAOA performance theorem; the latter is measured in the numerical experiments. If a trained state reaches probability $p_{\Good}$, standard amplitude amplification has the exact success formula
\begin{equation}
    p_r=
    \sin^2\!\left((2r+1)\arcsin\sqrt{p_{\Good}}\right),
    \label{eq:aa}
\end{equation}
with its own oracle cost. We therefore report the directly trained $p_{\Good}$ and do not include amplitude amplification in the principal result.

\textit{16-bit S-AES benchmark.---}
The numerical study uses public plaintexts selected from $0x6F6B$, $0x1234$, and $0xBEEF$ under the S-AES circuit in Eq.~(\ref{eq:saes-main}); the reversible module definitions and diagnostic-sandwich circuit specifications are given in the Supplement~\cite{Supplement}. The staged framework is
\[
    {\rm DDT}_{12}\rightarrow {\rm residual}_{3}\rightarrow {\rm GHD}_{3},
\]
when DDT data are available. For one public pair no DDT difference exists, so the first block is replaced by one-pair anchors such as residual/GHD-only, NPP-style Hamming-shell, or MIS-style Hamming-shell variants. The residual diagnostic maps are $\{\mathtt{inv\mbox{-}final\mbox{-}sbox},\mathtt{inv\mbox{-}round2},\mathtt{inv\mbox{-}two\mbox{-}rounds}\}$. With $m\ge2$ public pairs, the DDT block uses all $\binom{m}{2}$ pairwise differences.

\begin{center}
\refstepcounter{table}\label{tab:main-results}
\footnotesize\textbf{TABLE~\thetable.} Strengthened three-pair 16-bit S-AES instance. Final probabilities are obtained from the trained effective key-register circuit after Adam optimization of the variational angles. All runs use three public pairs, hence three pairwise DDT constraints and 12 DDT nibble components. The table reports ten runs from two independently seeded five-run batches; the true key is ranked first in every instance.

\begin{tabular}{@{}cccc@{}}
\hline
run & true key & final $p_{\Good}$ & largest false\\
\hline
1 & $0x652A$ & $0.370837$ & $0.004978$\\
2 & $0x1CC0$ & $0.521675$ & $0.009504$\\
3 & $0xC58D$ & $0.314952$ & $0.005563$\\
4 & $0xC8F9$ & $0.296658$ & $0.004484$\\
5 & $0xC7DB$ & $0.880889$ & $0.001719$\\
6 & $0x3832$ & $0.392327$ & $0.011056$\\
7 & $0xEA28$ & $0.315257$ & $0.007639$\\
8 & $0x2551$ & $0.627285$ & $0.005242$\\
9 & $0x6799$ & $0.275333$ & $0.002808$\\
10 & $0x4178$ & $0.682476$ & $0.007502$\\
\hline
mean & -- & $0.467769$ & $0.006050$\\
\hline
\end{tabular}
\end{center}

\begin{center}
\resizebox{0.98\columnwidth}{!}{%
\begin{tikzpicture}[
    font=\scriptsize,
    axis/.style={thin},
    bar2/.style={fill=orange!60, draw=orange!80!black},
    bar3/.style={fill=blue!45, draw=blue!70!black},
    ref2/.style={orange!80!black, dashed, thick},
    ref3/.style={blue!70!black, dashed, thick}
]
\draw[axis] (0,0) -- (10.9,0);
\draw[axis] (0,0) -- (0,10.35);
\foreach \y/\lab in {0/0,2/20,4/40,6/60,8/80,10/100} {
    \draw[axis] (-0.05,\y) -- (0,\y);
    \node[anchor=east] at (-0.08,\y) {\lab\%};
}
\draw[red!70!black, dashed] (0,1.0) -- (10.65,1.0);
\node[anchor=west, red!70!black] at (10.72,1.0) {10\%};

\draw[bar2] (0.20,0) rectangle (0.45,3.49874);
\draw[bar2] (1.20,0) rectangle (1.45,1.46937);
\draw[bar2] (2.20,0) rectangle (2.45,2.13070);
\draw[bar2] (3.20,0) rectangle (3.45,3.30207);
\draw[bar2] (4.20,0) rectangle (4.45,1.48376);
\draw[bar2] (5.20,0) rectangle (5.45,0.60368);
\draw[bar2] (6.20,0) rectangle (6.45,1.19251);
\draw[bar2] (7.20,0) rectangle (7.45,2.57993);
\draw[bar2] (8.20,0) rectangle (8.45,1.56273);
\draw[bar2] (9.20,0) rectangle (9.45,3.00464);

\draw[bar3] (0.55,0) rectangle (0.80,3.70837);
\draw[bar3] (1.55,0) rectangle (1.80,5.21675);
\draw[bar3] (2.55,0) rectangle (2.80,3.14952);
\draw[bar3] (3.55,0) rectangle (3.80,2.96658);
\draw[bar3] (4.55,0) rectangle (4.80,8.80889);
\draw[bar3] (5.55,0) rectangle (5.80,3.92327);
\draw[bar3] (6.55,0) rectangle (6.80,3.15257);
\draw[bar3] (7.55,0) rectangle (7.80,6.27285);
\draw[bar3] (8.55,0) rectangle (8.80,2.75333);
\draw[bar3] (9.55,0) rectangle (9.80,6.82476);

\draw[ref2] (0,2.08281) -- (10.45,2.08281);
\node[anchor=west, orange!80!black] at (10.52,2.08281) {mean $m=2$};
\draw[ref3] (0,4.67769) -- (10.45,4.67769);
\node[anchor=west, blue!70!black] at (10.52,4.67769) {mean $m=3$};

\foreach \x/\lab in {0.50/1,1.50/2,2.50/3,3.50/4,4.50/5,
                      5.50/6,6.50/7,7.50/8,8.50/9,9.50/10} {
    \node[anchor=north] at (\x,-0.08) {\lab};
}
\draw[bar2] (7.95,9.55) rectangle (8.20,9.80);
\node[anchor=west] at (8.28,9.675) {$m=2$};
\draw[bar3] (9.00,9.55) rectangle (9.25,9.80);
\node[anchor=west] at (9.33,9.675) {$m=3$};
\node[rotate=90, anchor=south] at (-0.58,5.10) {final $p_{\Good}$};
\node[anchor=north] at (5.05,-0.55) {matched random-key run};
\end{tikzpicture}%
}
\refstepcounter{figure}\label{fig:main-bars}
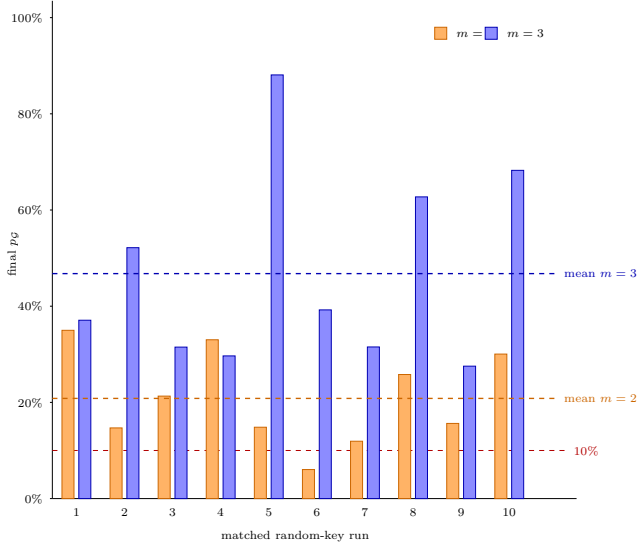
\footnotesize\textbf{FIG.~\thefigure.} Final public-consistent-key probability for the matched ten-instance comparison. Adjacent bars show the two-pair ($m=2$, orange) and three-pair ($m=3$, blue) conditions for the same random-key schedule. The red dashed line marks $10\%$; the colored dashed lines mark the corresponding means, $20.83\%$ and $46.78\%$.
\end{center}

Table~\ref{tab:main-results} and Fig.~\ref{fig:main-bars} report the three-pair benchmark and its matched two-pair control. The DDT stage contains three pairwise differences and 12 nibble components in the principal setting. The numerical campaign was not based on a single hand-tuned instance: two independently seeded five-run batches were executed, and the ten runs are reported as a fixed statistical sample to keep the presentation compact. No run was removed or selected on the basis of its outcome. Across the ten three-pair random-key instances, the mean final public-consistent-key probability is $46.78\%$, the minimum is $27.53\%$, and the largest observed false-key probability is $1.11\%$. Relative to the uniform 16-bit baseline $2^{-16}$, the mean is an enhancement of approximately $3.1\times10^4$. The correct key is ranked first in every instance. These data show that the staged phase design can convert public differential and inverse-round information into a pronounced key-register bias in the reduced cipher.

To quantify the dependence on the amount of public data, we repeated the same two-batch random-seed campaign with two public pairs while keeping the ten key/optimizer seed pairs fixed. Thus the comparison changes the number of pairwise DDT constraints but not the seed schedule or the optimization protocol. The two conditions comprise 20 completed random-seed runs in total, with ten matched instances reported for each condition. No run was discarded on the basis of its outcome. The corresponding aggregate statistics are
\begin{center}
\refstepcounter{table}\label{tab:pair-comparison}
\footnotesize\textbf{TABLE~\thetable.} Matched ten-instance comparison between two- and three-pair S-AES inputs. The reported probability is the independently evaluated public-consistent-key probability.

\resizebox{0.98\columnwidth}{!}{%
\begin{tabular}{@{}lccccc@{}}
\hline
public pairs & DDT components & mean $p_{\Good}$ & std. dev. & mean largest false & rank first\\
\hline
$m=2$ & $4$ & $0.208281$ & $0.092611$ & $0.017134$ & $10/10$\\
$m=3$ & $12$ & $0.467769$ & $0.193166$ & $0.006050$ & $10/10$\\
\hline
\end{tabular}
}
\end{center}
The three-pair mean is approximately $2.25$ times the two-pair mean, an absolute increase of $25.95$ percentage points. At the same time, the mean largest false-key probability decreases from $1.71\%$ to $0.60\%$. The result is consistent with the role assigned to the DDT stage: adding the third public pair supplies two additional pairwise differences and eight additional nibble components, improving separation before the residual stages.

\textit{Reversible realization and scope.---}
The 16-bit data use the exact induced action on the key register rather than a full gate-level simulation of all work registers. This is a reduced-model choice, not an assumption that a physical oracle can load an arbitrary phase table. Each phase can instead be implemented by reversible computation of the relevant DDT or residual diagnostic, a controlled phase based on a constant-size code, counter, or comparator, and uncomputation. The resulting key-register diagonal is the action of $U_f^\dagger P_f U_f$. Generic decompositions of that diagonal are used only as numerical consistency checks; they are not the proposed compilation of the physical oracle. A gate-level device would execute the reversible sandwiches directly and estimate the objective from measurements.

The Supplement~\cite{Supplement} gives the reversible circuit specifications, phase-module constructions, S-DES gate-level checks, numerical evaluation protocol, and circuit-resource scaling model. The direct-overlap calibration included there places large ciphertext-ground-state probabilities reported for small VQAA instances in the proper measurement context; it is not part of the structured 16-bit S-AES procedure.

\textit{Resources and limitations.---}
For direct sampling of the trained state, the expected number of repetitions needed to observe a public-consistent key is $O(1/p_{\Good})$. For a fixed schedule, the reversible gate cost is polynomial in the cipher size, number of public pairs, checkpoint count, and variational depths, whereas dense state-vector memory is exponential in the total logical width. With $m$ public pairs, all-pair DDT information gives $\binom{m}{2}$ pairwise constraints and four nibble components per pair in 16-bit S-AES; residual and GHD diagnostics scale as $m|\mathcal T|$. The DDT depth is an empirical variational parameter, not a theoretically optimal value. The present contribution is consequently a reduced-cipher mechanism study with an explicit reversible realization and a resource model, rather than a claim of asymptotic advantage over generic quantum search.

\textit{Concluding perspective.---}
The next step is a width-controlled sequence of gate-level AES-like experiments beginning with instances whose reversible workspaces exceed the practical dense state-vector regime. Such experiments should retain the public diagnostic construction, estimate training objectives from shots, verify only measured candidates, and report logical width, routed two-qubit depth, noise, shot cost, key rank, and $p_{\Good}$. Comparisons with Grover and other variational baselines should use the same oracle and measurement conventions. Within its stated scope, the present study shows that DDT preparation followed by inverse-round residual refinement is a reproducible key-concentration mechanism for S-AES and supplies a concrete circuit hypothesis for larger quantum experiments.

\textit{Data and code availability.---}
The numerical data supporting the figures and tables are generated by the reduced-cipher simulations described in the Supplement. The data, run logs, and implementation details are available from the corresponding author upon reasonable request. A public repository will be released upon acceptance.

\textit{Acknowledgments.---}
This work was supported by the National Natural Science Foundation of China (Grant Nos. 61871120 and 62071240), the Natural Science Foundation of Jiangsu Province (Grant Nos. BK20191259 and BK20220804), the Innovation Program for Quantum Science and Technology (Grant No. 2021ZD0302901), the Six Talent Peaks Project of Jiangsu Province (Grant No. XYDXX-003), and the Jiangsu Funding Program for Excellent Postdoctoral Talent (Grant No. 2022ZB107).

\bibliography{prl_qaoa_aes_refs}

\end{document}